%% file: main.tex
\documentclass[sigconf]{acmart}

\usepackage{bm}
\usepackage{booktabs}
\usepackage{bbding}
\usepackage{array}
\usepackage{arydshln}
\usepackage{lipsum}
\usepackage{threeparttable}
\usepackage{enumerate}
\usepackage{float}
\usepackage{placeins}
\usepackage{subcaption}  

\definecolor{cvprblue}{rgb}{0.21,0.49,0.74}
\acmConference[Conference acronym 'XX]{Make sure to enter the correct
  conference title from your rights confirmation emai}{June 03--05,
  2018}{Woodstock, NY}
\acmISBN{978-1-4503-XXXX-X/18/06}

\title{Behavioral Feature Boosting via Substitute Relationships for E-commerce Search}

\begin{document}
\input{authors}
\input{abstract}
\input{ccsxml}
\keywords{Cold-start, Behavioral Feature, Substitute, E-commerce, Search}
\maketitle

\input{main_body}


\bibliographystyle{ACM-Reference-Format}
\bibliography{reference}


\end{document}

%% file: authors.tex
\author{Chaosheng Dong, Michinari Momma, Yijia Wang, Yan Gao and Yi Sun}
\email{{chaosd,michi,yijiawan,yanngao,yisun}@amazon.com}
\affiliation{%
  \institution{Amazon}
  \city{Seattle}
  \state{Washington}
  \country{USA}
}








%% file: abstract.tex
\begin{abstract}
On E-commerce platforms, new products often suffer from the cold-start problem: limited interaction data reduces their search visibility and hurts relevance ranking. To address this, we propose a simple yet effective \emph{behavior feature boosting} method that leverages substitute relationships among products (BFS). BFS identifies substitutes—products that satisfy similar user needs—and aggregates their behavioral signals (e.g., clicks, add-to-carts, purchases, and ratings) to provide a warm start for new items. Incorporating these enriched signals into ranking models mitigates cold-start effects and improves relevance and competitiveness. Experiments on a large E-commerce platform, both offline and online, show that BFS significantly improves search relevance and product discovery for cold-start products. BFS is scalable and practical, improving user experience while increasing exposure for newly launched items in E-commerce search. The BFS-enhanced ranking model has been \textbf{launched} in production and has served  customers since 2025.
\end{abstract}

%% file: ccsxml.tex
\ccsdesc[500]{Information systems~Information retrieval~Retrieval models and ranking~Learning to rank}

%% file: main_body.tex
\section{Introduction}
On E-commerce platforms, search relevance systems play a critical role in helping users efficiently discover products. However, new products often face the \textit{cold-start problem} \cite{schein2002methods,lika2014facing,gope2017survey,ko2022survey}: limited user interaction data makes it difficult for them to rank well in search results. As a result, these products receive low visibility, delaying discovery and potentially reducing sales. The cold-start problem is especially acute in competitive markets, where visibility and discoverability directly affect customer satisfaction and sales performance.

Traditional E-commerce search rankers rely heavily on behavioral features---such as click-through rate (CTR), add-to-carts, purchases, and review ratings---derived from historical customer interactions \cite{liu2009learning,agichtein2006improving}. For new products, the lack of such interactions leads to suboptimal ranking decisions. Because these signals are central to estimating relevance and user satisfaction, cold-start items are often pushed to lower positions, making it difficult to attract initial customer attention.

To address this challenge, we propose \textit{Behavior Feature Boosting} through substitute relationships (BFS). Our approach identifies substitute products \cite{mcauley2015inferring,zheng2021heterogeneous,jian2022multi,zhou2023multi}---items that satisfy similar user needs---and transfers (aggregates) their behavioral signals to new products. Although substitutes are not identical, they can fulfill similar purposes for users; therefore, their historical interactions can provide a useful proxy signal for cold-start items. By leveraging substitutes, BFS supplies new products with stronger behavioral features and improves their visibility in search.

In this work, we focus on an important behavioral feature in E-commerce: \textit{Sales Velocity (SV)}. SV captures decayed order counts over a recent time window, so more recent purchases contribute more heavily and the feature reflects current popularity trends. We boost SV for cold-start products by aggregating SV from their substitutes. We consider simple summary statistics (e.g., mean and maximum) as well as attention-based aggregation strategies. The resulting aggregated values are then used as enhanced behavioral signals, making new products more competitive in search rankings.

\subsection{Related Work}

BFS is related to collaborative filtering (CF), a widely used technique in recommendation systems \cite{koren2021advances,aljunid2025collaborative}. CF methods, including user-based and item-based approaches, leverage historical user--item interactions to identify patterns and generate recommendations by finding similarities among users or items \cite{sarwar2001item,guan2024hybrid}. Although CF is effective for personalized recommendation, it primarily operates at the entity level, modeling relationships among specific users and items. As a result, it typically depends on large interaction matrices and expensive similarity computations, and it remains ill-suited to cold-start settings where interaction data are sparse or missing \cite{schein2002methods}.
In contrast, BFS takes a feature-level perspective by transferring behavioral signals from substitute products to new items. This design bypasses the need for direct user--item interaction histories by using precomputed substitute relationships to propagate behavioral information.
This feature-level focus positions BFS as a complementary approach to CF, directly addressing CF's limitations in environments where product cold start is a major challenge.

\begin{figure*}
    \centering
    \includegraphics[width=1\linewidth]{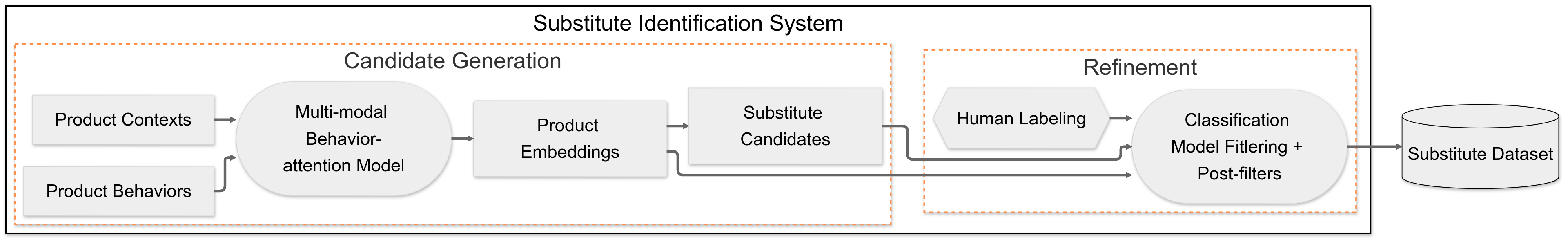}
    \caption{Diagram of the substitute identification system.}
    \label{fig:substitute-identification-system}
\end{figure*}

Our key contributions are summarized as follows:

\begin{itemize}
    \item We introduce \textit{Behavioral Feature Boosting} (BFS), a method for mitigating the cold-start problem in E-commerce search by transferring behavioral signals (e.g., Sales Velocity) from substitute products to new products, thereby improving their visibility and relevance in search rankings.
    \item We validate BFS through offline experiments and online A/B tests on a large E-commerce platform, demonstrating substantial improvements in key metrics such as GMV, Sales Units, and new-product discoverability. The BFS-enhanced ranking model has been \textbf{launched} in production and has served customers since 2025.
\end{itemize}

\section{Methodology}

In this section, we describe our approach to Behavior Feature Boosting through substitute relationships (BFS). The methodology comprises three main components: substitute identification, behavioral feature aggregation, and integration into the search ranking model.

\subsection{Definition of Behavioral Feature}

A behavioral feature in E-commerce search refers to a measurable signal derived from user interactions with products, such as clicks, add-to-carts, purchases, ratings, and other engagements. Behavioral features capture patterns of user behavior and often reflect an item's popularity, relevance, or attractiveness based on historical data. In contrast, a content-based feature is derived from a product's intrinsic attributes or metadata, such as the title, description, category, images, and specifications.

Let \( U \) be the set of users, \( P \) be the set of products, \( A \) be the set of possible actions (e.g., clicks, purchases, views), \( T \) be the set of timestamps representing the time of interaction. A behavioral feature \( f \) is a function:
\begin{equation}
    f: U \times P \times A \times T \rightarrow \mathbb{R} \label{eq:function_mapping},
\end{equation}
where \( \mathbb{R} \) is the set of numbers representing the feature value (e.g., count, probability, rate) that quantifies the interaction or behavior.




\subsection{Substitute Identification}

Identifying high-quality substitutes is critical to BFS. We define substitutes as products in the same category that fulfill similar functions. In practice, substitutes can be identified using:
\begin{itemize}
    \item \textbf{Product Attributes:} Category, brand, specifications, etc.
    \item \textbf{Content-Based Features:} Title, description, etc.
    \item \textbf{User Behavior:} Co-clicks, view-to-purchases, etc.
\end{itemize}

As shown in Figure \ref{fig:substitute-identification-system}, the substitute identification framework has two main components: (a) \textbf{Candidate generation.} We train unified multimodal product embeddings from images and text using substitute relations, targeting high recall at a fixed precision. Concretely, we add a behavior-attention transformer on top of BLIP2 \cite{li2023blip} to fine-tune product embeddings. We then build an embedding index based on similarity. At serving time, multiple queries are executed in parallel to efficiently retrieve the top-$K$ most similar items. (b) \textbf{Classification-based filtering.} Because nearest-neighbor retrieval can produce low-precision candidates, we apply a classification model to refine the results. The classifier takes as input the query-product embedding, the candidate-substitute embedding, and their embedding difference. Training labels are obtained via human auditing, indicating whether a product pair is a substitute. When the target precision cannot be met for certain categories, we add post-filters on specific attributes (e.g., color and size) to further improve precision.

\begin{figure}[h]
    \centering
    \includegraphics[width=1\linewidth]{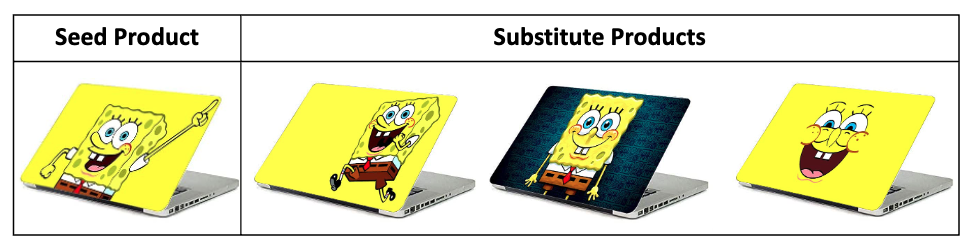}
    \caption{Example substitute products. The seed item is a cartoon laptop-skin sticker.}
    \label{fig:substitute-example}
\end{figure}

Our substitute identification system is designed to function even when only a subset of the mentioned data sources is available. We show an example of substitute products in Figure \ref{fig:substitute-example}.

\subsection{Behavioral Feature Boosting Design}
BFS mitigates the cold-start problem for new products in E-commerce search by leveraging behavioral signals from substitute products that satisfy similar user needs. By aggregating behavioral features from substitutes, new products receive a ``warm start'' signal that improves their visibility and discoverability in search.

\textbf{Formal Definition:}  
Let \( p \) be a new product, and \( S_p = \{s_1, s_2, \ldots, s_n\} \) be the set of substitute products for \( p \). Let \( f_b \) be a behavioral feature (e.g., clicks, purchases) associated with a product. The boosted behavioral feature \( f_b^{*}(p) \) for the product \( p \) is defined as:
\begin{equation}
    f_b^{\text{Subs}}(p) = g(f_b(s_1), f_b(s_2), \ldots, f_b(s_n)),
    \label{eq:substitution_function}
\end{equation}
where \( g \) is an aggregation function such as the mean, maximum, or 75th percentile of the behavioral features of the substitutes.

\textbf{Example Aggregation Strategies:}
\begin{itemize}
    \item \textbf{Mean Aggregation:} $ \frac{1}{n} \sum_{i=1}^{n} f_b(s_i) $,
    \item \textbf{Max Aggregation:} $  \max_{i=1}^{n} f_b(s_i) $,
    \item \textbf{Attention-based Aggregation:} $ \frac{1}{\sum_{i=1}^{n} \langle p,s_i\rangle} \sum_{i=1}^{n} \langle p,s_i\rangle f_b(s_i) $.
\end{itemize}


\subsection{Integration BFS into Ranking Model}

We incorporate $f_b^{\text{Subs}}(p)$ as an additional feature in the search ranking model. The ranking score for a product \(p\) given a query $q$ is defined as:
\begin{equation}
    \text{Score}(q,p) = \pi_q(\mathbf{X}, f_b^{\text{Subs}}(p)),
\end{equation}
where \( \mathbf{X} \) represents other features (e.g., textual relevance, product attributes), and \( \pi_q \) is the ranking model (e.g., LambdaMART \cite{burges2010ranknet} or ListNet \cite{cao2007learning,qin2021neural}).
$f_b^{\text{Subs}}(p)$ is refreshed periodically (e.g., every four hours) to ensure it reflects the latest behavioral data.


\textbf{Key Steps in BFS:}
\begin{enumerate}
    \item \textbf{Identify Substitutes:} Find a set of substitutes \( S_p \) for the new product \( p \) based on criteria like product attributes, user behavior, or content similarity.
    \item \textbf{Aggregate Behavioral Features:} Compute the behavioral feature \( f_b^{\text{Subs}}(p) \) by aggregating the behavioral signals \( f_b(s_i) \) of the substitutes \( s_i \) using an appropriate function \( g \).
    \item \textbf{Integrate into the ranking model:} Use \( f_b^{\text{Subs}}(p) \) as an input feature to boost the relevance score, particularly for products with limited customer interactions.
\end{enumerate}

\section{Experiments}

We conducted experiments on a large E-commerce platform in 2024 to evaluate the effectiveness of BFS in mitigating the cold-start problem and improving search relevance.

\subsection{Experimental Setup}
This subsection describes the experimental setup, including the dataset, product substitutes, models, and evaluation metrics.

\textbf{Dataset:} We used a dataset of historical search and purchase logs collected over a one-month period. The dataset contains millions of interactions spanning a wide range of categories (e.g., electronics, home improvement, and automotive). 

\textbf{Behavioral feature:} A key feature in our E-commerce ranking model is a time-decayed measure of product popularity known as \textit{Sales Velocity} (SV). Intuitively, SV can be viewed as a (possibly weighted) sum of purchases for a product over a recent fixed-length time window. In practice, SV includes refinements in which each unit sale is gradually decayed to obtain a smoother popularity signal (e.g., by blending multiple half-life decay functions).






\textbf{Product substitutes:} The dataset includes both new and established products. For each product, we identify a set of substitutes \( S_p = \{s_1, s_2, \ldots, s_n\} \) based on similarity in user behavior and product attributes. Substitute relationships are precomputed and stored in a lookup table for efficient retrieval during model inference. On average, each new product has 5 substitutes, with a maximum of 10.

\begin{figure}[h]
    \centering
    \includegraphics[width=0.99\linewidth]{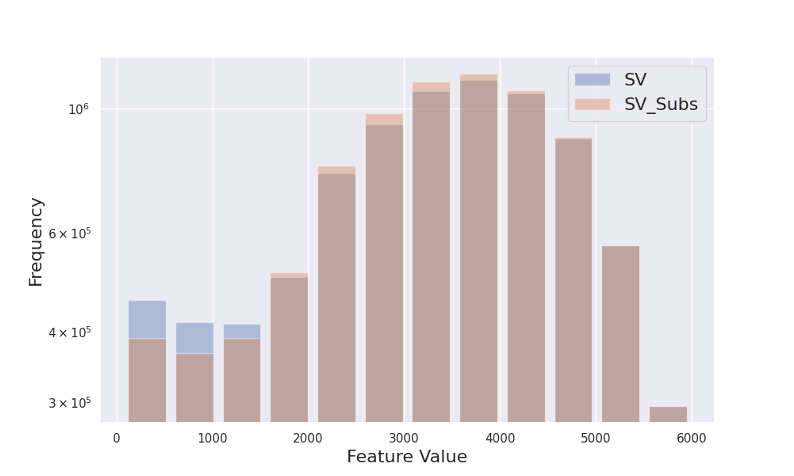}
    \caption{Distribution of SV and SV\_Subs. Blue denotes SV, orange denotes SV\_Subs, and brown indicates the overlap between the two histograms.}
    \label{fig:histogram}
\end{figure}

\textbf{SV\_Subs:} To avoid underestimating already popular products, we apply the adjustment
$f_b^{\text{Subs}}(p) \leftarrow \max\{f_b^{\text{Subs}}(p), f_b(p)\}$.
This ensures that the ranking/scoring system does not penalize strong-performing products due to substitution effects or model artifacts. In Figure \ref{fig:histogram}, SV\_Subs has lower mass in the low-value range (0--1000) and higher mass in the mid-value range ($\sim$2000--4000) compared with SV, while the two features behave similarly in the high-value range (4000--6000). This pattern indicates that SV\_Subs successfully boosts products with relatively low sales.

\textbf{Models:} We trained two LambdaMART \cite{burges2010ranknet,Mahapatra_2023,Dong_2025} learning-to-rank models on 3M queries sampled from a one-month window, using the same loss function and optimization procedure. Specifically,
\begin{itemize}
    \item \textbf{Baseline (T1):} The baseline search ranking model trained with the new dataset without SV\_Subs feature.
    \item \textbf{T2:} The model trained with the new dataset augmented with SV\_Subs using the mean aggregation strategy.
\end{itemize}

\begin{figure}
    \centering
    \includegraphics[width=0.8\linewidth]{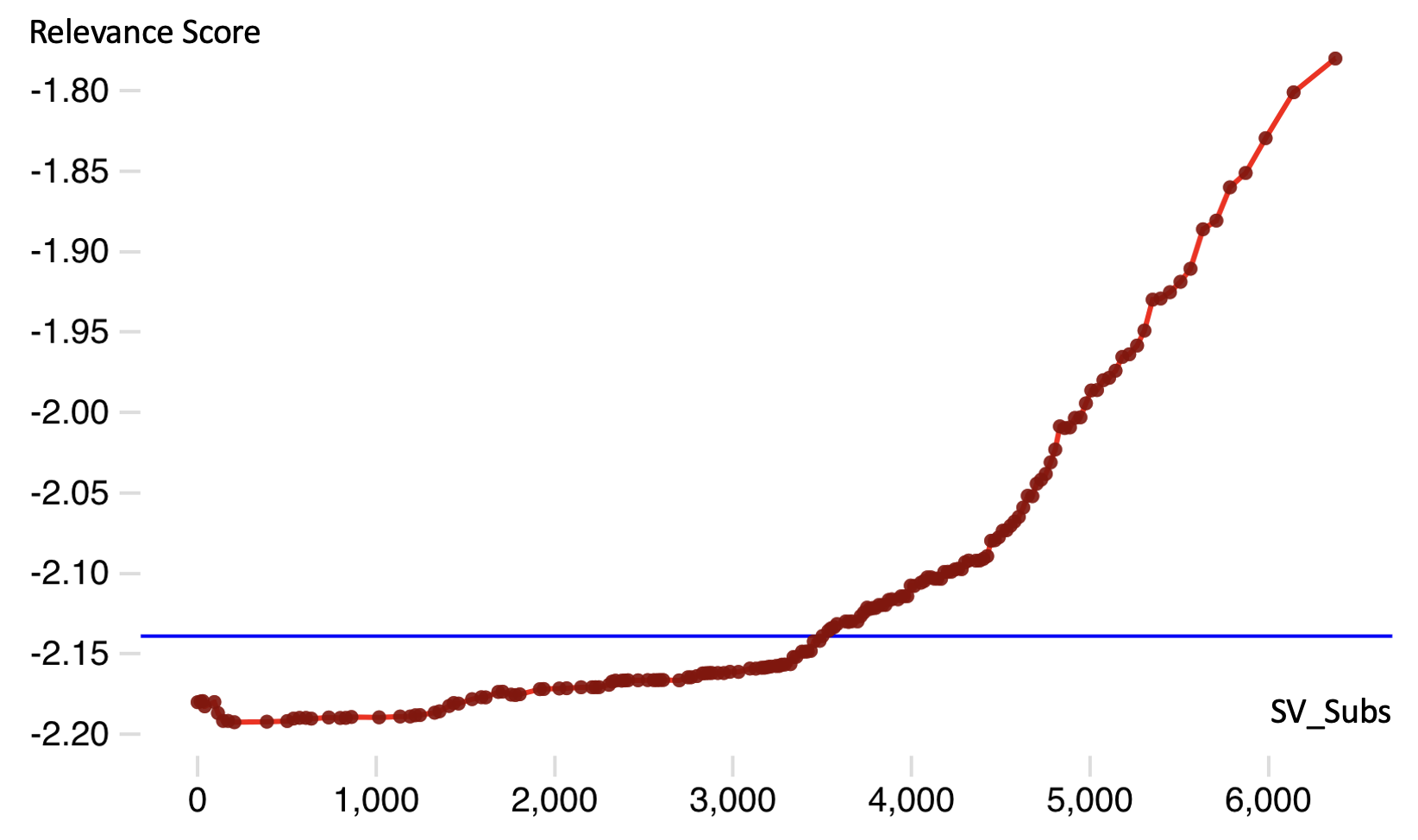}
    \caption{Relevance score as a function of the feature value.}
    \label{fig:relevance_change}
\end{figure}

Figure \ref{fig:relevance_change} shows how the relevance score for T2 changes with SV\_Subs. Because the ranker orders products by this score, higher values correspond to more prominent placements in the search results. The curve exhibits a smooth upward trend, indicating that SV\_Subs contributes positively to ranking. Notably, the marginal effect increases with the feature value: products with SV\_Subs below 3600 are demoted and receive relevance scores well below the average (blue line). Moreover, the contribution of SV\_Subs continues to increase without clear saturation.

\textbf{Evaluation Metrics:} We report the following key metrics:

\begin{itemize}
    \item \textbf{GMV:} The total sales (in dollars) over a given period on the e-commerce platform.
    \item \textbf{New Product GMV (NP-GMV):} GMV attributed to newly launched products within a given time frame.
    \item \textbf{New Product Sales Units (NP-Sales Unit):} The total number of units sold for newly launched products.
\end{itemize}

\subsection{Offline Evaluation}

\begin{table}[h]
\centering
\caption{Gains in NDCG@10, wrt the production ranker.}
\label{tab: offline_results}
\small 
\setlength{\tabcolsep}{4pt} 
\renewcommand{\arraystretch}{0.9} 
\resizebox{0.5\textwidth}{!}{ 
\begin{tabular}{lcccc}
\toprule
\textbf{Model} & \textbf{Relevance} & \textbf{Purchases} & \textbf{Sales Units} & \textbf{Clicks} \\
\midrule
Baseline (T1)   & -0.05\%    & +0.67\%    & +0.28\%    & +1.28\% \\
T2 (Mean)       & -0.08\%    & +1.32\%    & +0.51\%    & +1.82\%  \\
T3 (Max)       & +0.21\%    & +0.79\%    & +0.51\%    & +0.32\% \\
T4 (Attention)        & +0.60\%    & +0.96\%    & +0.83\%    & +0.12\%   \\
\bottomrule
\end{tabular}
}
\end{table}

Table \ref{tab: offline_results} summarizes the impact of SV\_Subs in the offline evaluation. Models augmented with SV\_Subs (T2) achieve the largest gains in key customer-behavior metrics: Purchases (in dollars), Sales Units, and Clicks all improve relative to both the production ranker and the baseline model (T1, refreshed production ranker). Overall, these results suggest that BFS can improve product discoverability and drive sales. However, offline evaluation in our setting can be sensitive to counterfactual bias; therefore, we additionally conducted an online A/B test to obtain an unbiased estimate of the treatment effect.

    

    

    

\begin{figure}
    \centering
    \includegraphics[width=0.9\linewidth]{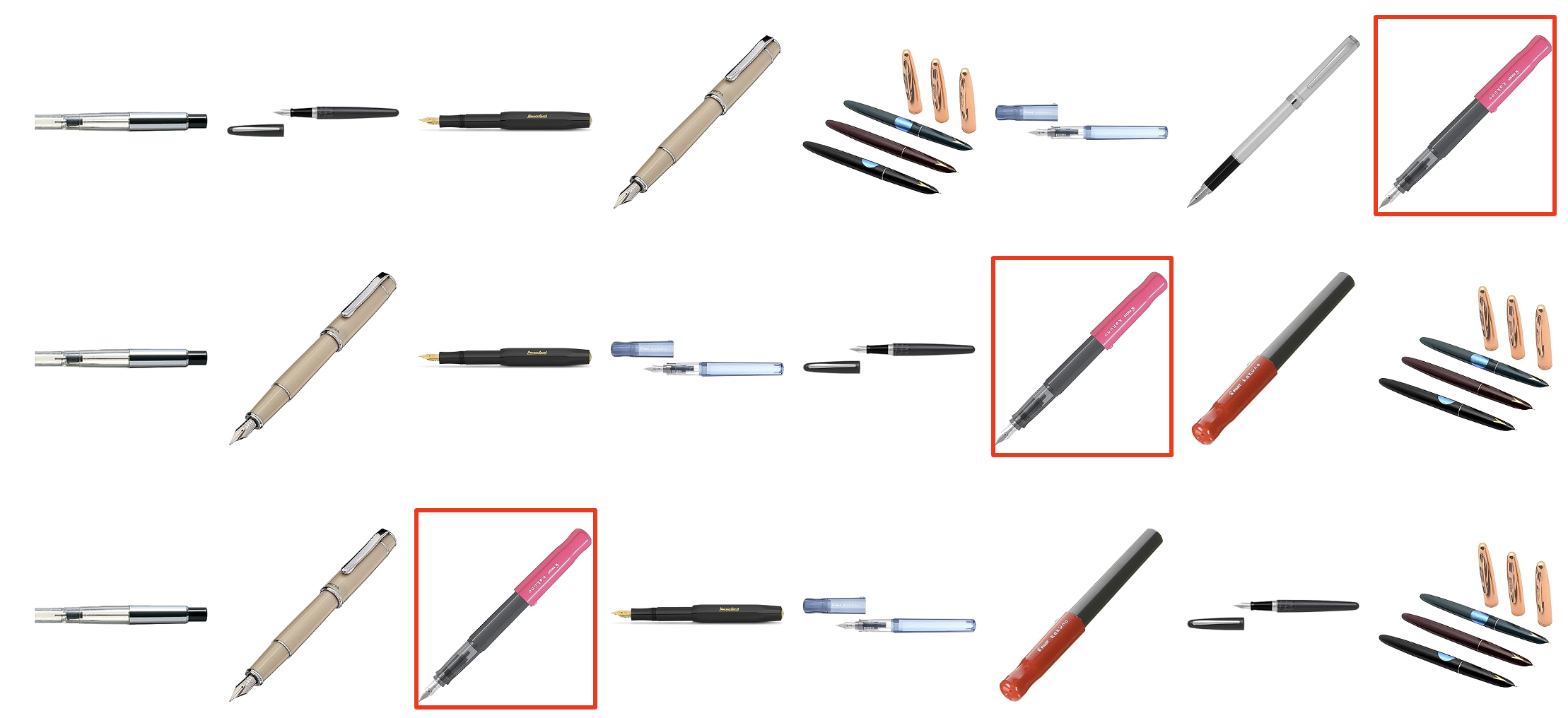}
    \caption{Top-8 search results for the query \textit{pilot explorer fountain pen} during the online A/B test. The first, second, and third rows show the ranked lists produced by the production ranker, T1, and T2, respectively.}
    \label{fig:qualatative}
\end{figure}

\subsection{Online Evaluation on a large E-commerce platform}
We conducted a four-week A/B test on a large E-commerce platform comparing T1 and T2, motivated by its strong offline performance. The results show that T2 outperforms both the production ranker and T1 across all metrics, demonstrating its ability to drive sales growth. Specifically, T2 improves GMV by +0.11\%, Sales Units by +0.22\%, NP-GMV by +0.18\%, and NP-Sales Units by +0.35\%, indicating consistent gains in both monetary and volume measures. The largest improvement is in NP-Sales Units, highlighting T2's effectiveness in increasing engagement with new products. In contrast, the baseline model T1 shows small declines across metrics. Overall, these online results confirm that the proposed features help surface under-discovered products, leading to higher engagement and more purchases. Following the experiment, we launched T2 in production, where it has been serving a large E-commerce platform customers since 2025.

\begin{table}
\centering
\caption{Online impact of SV\_Subs on search metrics. We report percentage gains relative to the production ranker; statistical significance is denoted by * (Probability of Positive Return $\geq$ 0.66).}
\label{tab: online_results}
\small 
\setlength{\tabcolsep}{5pt} 
\renewcommand{\arraystretch}{0.9} 
\resizebox{0.5\textwidth}{!}{ 
\begin{tabular}{lcccc}
\toprule
\textbf{Model}       & \textbf{GMV} & \textbf{Sales Units} & \textbf{NP-GMV} & \textbf{NP-Sales Units} \\
\midrule
Baseline (T1)        & -0.19\%    & -0.10\%    & -0.26\%    & -0.11\% \\
T2 (Mean)            & $+0.11\%^*$    & $+0.22\%^*$     & $+0.18\%^* $    & $+0.35\%^*$  \\
\bottomrule
\end{tabular}
}
\end{table}



\subsection{Qualitative Analysis}
In Figure \ref{fig:qualatative}, we present the top eight search results rendered in 2024 during A/B testing to illustrate how incorporating SV\_Subs into the ranking qualitatively impacts the customer experience. For the query \textit{pilot explorer fountain pen,} a pink Pilot Fountain Pen Medium was initially ranked eighth by the production ranker. This product had an SV of 89 in August, which is relatively low, leading to its placement in the eighth position despite having one of the highest customer ratings among the listed fountain pens. T1 made a slight adjustment, moving it to sixth. However, T2 assigned it an SV\_Subs value of 297, significantly higher than the original SV, pushing it up to third. This example demonstrates the effectiveness of leveraging the BFS signal to enhance the visibility of under-discovered products, ultimately improving the customer search experience.



\section{Conclusion}
In this paper, we presented Behavior Feature Boosting through substitute relationships (BFS) to mitigate the cold-start problem in E-commerce search. By leveraging behavioral signals from established substitute products, BFS provides new products with a ``warm start,'' improving their discoverability in search rankings. We instantiate BFS using Sales Velocity (SV), where SV is aggregated over substitutes to provide an enriched sales signal for new products and to improve their visibility and competitiveness in search results. Through offline experiments and real-world online A/B tests on a large E-commerce platform, we demonstrated that BFS improves key business metrics, including GMV, Sales Units, and new-product discoverability. Overall, the results highlight the scalability and practicality of BFS for large-scale e-commerce platforms. Future work will explore dynamic substitute identification, additional behavioral signals, and more advanced aggregation strategies (e.g., attention-based methods).

\section*{Short Bio of the Main Presenter}
Dr. Chaosheng Dong is a machine learning scientist specializing in search relevance, generative information retrieval, and multi-objective optimization. He is currently a Tech Lead at Amazon, where he has helped transition Amazon Search from a single-task to a multi-task system and improve key search performance metrics. Dr. Dong received his Ph.D.\ in Operations Research from the University of Pittsburgh and has published in top-tier conferences including ICML, KDD, and NeurIPS. His research interests include multi-task learning, ranking models, and recommendation systems.